\documentclass[prb,twocolumn,showpacs,amsmath,amssymb,floatfix]{revtex4}
\usepackage{color,graphics,epsfig,rotating}
\usepackage{graphicx}% Include figure files
\usepackage{dcolumn}% Align table columns on decimal point
\usepackage{bm}% bold math
\usepackage{subfigure}

\begin{document}

\title{Electron-phonon superconductivity and charge density wave
  instability in the layered titanium-based pnictide
  BaTi$_2$Sb$_2$O}

\author{Alaska Subedi}
\affiliation{Centre de Physique Th\'eorique, \'Ecole Polytechnique,
  CNRS, 91128 Palaiseau Cedex, France}

\date{\today}

\begin{abstract}
  I present the results of first principles calculations of the phonon
  dispersions and electron-phonon coupling for BaTi$_2$Sb$_2$O. The
  phonon dispersions show a weak lattice instability near the zone
  corners that leads to a charge-density wave phase. The calculations
  of the electron-phonon coupling reveal strong coupling, especially
  to the in-plane Ti modes. The total coupling is large enough to
  readily explain the superconductivity in this compound. As the Fermi
  surfaces are disconnected with different orbital character weights,
  this compound is likely to host a multiband superconductivity.
\end{abstract}


\maketitle

\section{Introduction}

The discovery of high-temperature superconductivity in iron based
compounds by Kamihara \textit{et al.}\cite{kami08} has increased
interest and intensified efforts to find superconductors in new
families of compounds. These efforts have focused on searching for
materials with structural, electronic and magnetic properties similar
to those of the iron or cuprate superconductors. In particular, the
ideal candidates are thought to be materials with a layered structure
that are in proximity to a spin-density wave (SDW) instability due to
Fermi surface nesting like in the iron based superconductors or a Mott
insulating phase due to strong on-site Coulomb repulsion as in the
cuprates.

Recently, Yajima \textit{et al.}\ have reported superconductivity in
BaTi$_2$Sb$_2$O with a $T_c$ of 1.2 K.\cite{yaji12} Superconductivity
in Ba$_{1-x}$Na$_x$Ti$_2$Sb$_2$O with a maximum $T_c$ of 5.5 K has
also been reported by Doan \textit{et al.}\cite{doan12} In addition to
the superconducting transition, there is also an anomaly at $T_a$ = 54
K that manifests by showing strong features in the measurements of
susceptibility, resistivity, and specific heat. The microscopic
mechanism and the order parameter for this transition have not been
determined but been ascribed to either a charge-density wave (CDW) or
SDW instability. Furthermore, the superconducting phase coexists with
this CDW or SDW phase.

This compound occurs in a tetragonal structure with space group
$P4/mmm$ and consists of square planes of transition element Ti. The O
atoms are placed alternately at the center of the Ti squares such that
each Ti atom has two O nearest neighbors. The Sb atoms are placed
above and below the center of the Ti squares that do not contain O
atoms. The Ti$_2$Sb$_2$O layers are stacked alternately with layers of
Ba atoms along the $c$ axis. This layered structure is similar to the
iron based or cuprate superconductors as all of them have square
planes of transition-metal atoms as a structural motif.

In addition, Singh has further highlighted similarities in the
electronic and magnetic properties between BaTi$_2$Sb$_2$O and the
iron superconductors based on first principles
calculations.\cite{sing12} The Fermi surface shows substantial nesting
and leads to an antiferromagnetic instability as in the case of iron
superconductors. This nested Fermi surface is consistent with the
results of Pickett on a similar compound Na$_2$Sb$_2$Ti$_2$O, although
he was unable to find a magnetically ordered state.\cite{pick98} The
electronic structure of Na$_2$Ti$_2$Sb$_2$O was also studied by de
Biani \textit{et al.}\ using tight-binding calculations, and they
also find substantial nesting in the Fermi surface that may lead to a
CDW instability.\cite{debi98}

In any case, the presence of a nested Fermi surface and
antiferromagnetic instability in BaTi$_2$Sb$_2$O leads Singh to
predict a spin-fluctuation mediated superconductivity similar to that
of the iron based superconductors.\cite{mazi08} The pairing state is
predicted to have a sign-changing $s$-wave symmetry that is different
from that of the iron based superconductors. Furthermore, spin
fluctuations are repulsive in the singlet channel and hence are
strongly pair-breaking for the attractive electron-phonon
interaction. Thus, it is reasonable to expect that the
superconductivity in BaTi$_2$Sb$_2$O is mediated by spin fluctuations,
especially if the phase transition at $T_a$ = 54 K that coexists with
the superconducting phase is due to SDW instability.

However, there are some differences between BaTi$_2$Sb$_2$O and other
superconductors near magnetism. The maximum $T_c$ = 5.5 K of this
compound is significantly smaller than that of the iron based or
cuprate superconductors. This could be due to a small value of the
Stoner parameter for Ti\cite{ande85} or pair-breaking effects of the
electron-phonon interaction. It is worthwhile to note that
Sr$_2$RuO$_4$, which is purported to be an unconventional
superconductor,\cite{rice95} also has a small $T_c$ of 1
K.\cite{maen94} A more glaring difference is the very sensitive
dependence of superconductivity on other constituent elements of the
compound. A transition similar to the one in BaTi$_2$Sb$_2$O is also
seen in other members of the family that includes
Na$_2$Ti$_2$$Pn$$_2$O ($Pn$ = As, Sb), BaTi$_2$As$_2$O,
(SrF)$_2$Ti$_2$$Pn$$_2$O ($Pn$ = As, Sb), and
(SmO)$_2$Ti$_2$Sb$_2$O,\cite{adam90,axte97,ozaw00,ozaw01,ozaw04,liu09,wang10,liu10,
  ozaw08} but superconductivity has so far only been observed in Na
doped BaTi$_2$Sb$_2$O. This is in contrast to iron superconductors
which exhibit superconductivity for a wide variety of cations and
pnictogens. In the cuprates, superconductivity is also found for a
wide variety of interlayer fillings that separate the CuO$_2$
layers. Even in the ruthenates, doping at the cation site or changing
the interlayer spacing reveals various magnetic interactions. This
suggests that the magnetism and mechanism for superconductivity in
doped BaTi$_2$Sb$_2$O are quite different from those of other
unconventional superconductors, and perhaps lattice instabilities and
electron-phonon coupling play important roles.

In this paper, I present the results of first principles calculations
that show presence of a CDW instability and a total electron-phonon
coupling strong enough to yield a conventional superconductivity in
BaTi$_2$Sb$_2$O. The phonon dispersions reveal a weak lattice
instability near the Brillouin zone corners. This instability is
associated with the elongation or compression of the Ti squares
without an enclosed O such that the Ti squares with an O rotate either
clockwise or counterclockwise by a small amount. I also find presence
of strong electron-phonon couplings, especially near the zone
corners. The total electron-phonon coupling $\lambda_{\textrm{ep}}$ =
1.28 for the undistorted structure or $\lambda_{\textrm{ep}}$ = 0.55
for the distorted structure is large enough to readily explain the
superconductivity in this material.

\section{Methods}
The phonon dispersions and electron-phonon results presented here were
obtained using density-functional perturbation theory\cite{dfpt}
within the generalized gradient approximation of Perdew, Burke, and
Ernzerhof\cite{pbe} as implemented in the Quantum-ESSPRESSO
package.\cite{qe} I used the experimental lattice parameters ($a$ =
4.11039 \AA and c = 8.0864 \AA)\cite{yaji12} but relaxed the internal
Sb height parameter $z_{\textrm{Sb}}$. I obtained a value for
$z_{\textrm{Sb}}$ = 0.2493, which agrees well with both the
experimental\cite{yaji12,doan12} and calculated\cite{sing12}
values. This is in contrast to the underestimation from density
functional calculations found in the iron based
superconductors.\cite{mazi08} This indicates that magnetism and/or
lattice dynamics in BaTi$_2$Sb$_2$O might be different from those of
the iron based superconductors.

I used pseudopotentials that were generated with the electronic
configurations 5$s^2$5$p^6$5$d^0$6$s^2$6$p^0$,
3$s^2$3$p^2$4$s^2$3$d^1$, 5$s^2$5$p^3$5$d^0$, and
2$s^2$2$p^4$3$d^{-2}$, respectively, for Ba, Ti, Sb, and O. In
particular, it was necessary to include the semicore states to get the
electronic structure in close agreement with the results of the
full-potential calculations. I used the cut-offs of 60 and 600 Ry
for basis-set and charge density expansions, respectively. An
$8\times8\times4$ $\mathbf{k}$-grid was used for the Brillouin zone
integration during the self-consistency. For the undistorted
structure, the dynamical matrices were calculated on an
$8\times8\times4$ $\mathbf{q}$-grid, while the double-delta
integration in the calculation of the electron-phonon spectral
function was done on a $24\times24\times12$ grid. For the distorted
structure, I calculated the dynamical matrices on a $4\times4\times4$
$\mathbf{q}$-grid and used $18\times18\times8$ grid for the
double-delta integration. I used the generalized full-potential method
as implemented in the WIEN2k package\cite{wien2k} to perform the
relaxation of the internal coordinates of a
$(\sqrt{2}\times\sqrt{2}\times1)$ supercell. The muffin-tin radii of
2.4, 1.7, 2.4, and 1.7 Bohr were used for Ba, Ti, Sb, and O,
respectively, and an $8\times8\times8$ $\mathbf{k}$-grid was used for
the Brillouin zone integration.

\begin{figure} %% [tbp]
  \includegraphics[width=\columnwidth]{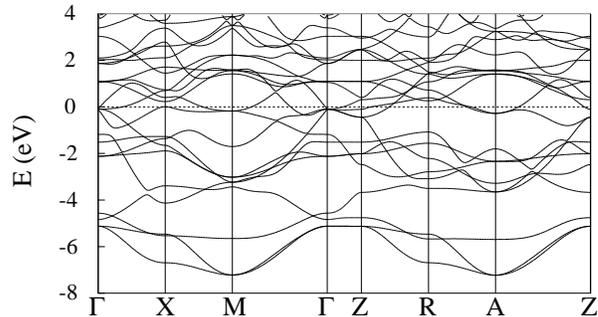}\\
  \caption{Calculated GGA band structure of non-spin-polarized
    BaTi$_2$Sb$_2$O in the undistorted structure with
    experimental lattice parameters but relaxed internal
    coordinates.}
  \label{fig:bnd}
\end{figure}

\section{Electronic structure}
The electronic structure and magnetism of BaTi$_2$Sb$_2$O have been
well described by Singh.\cite{sing12} I summarize his results here for
the sake of completeness. The band structure of this compound
calculated using the pseudopotentials and plane-wave basis set is
shown in Fig.~\ref{fig:bnd}. This agrees reasonably well with Singh's
full-potential band structure that also includes the spin-orbit
interaction. I also did full-potential band structure calculations
without the spin-orbit coupling, and I find excellent agreement with
the results from pseudopotential calculation. This shows that the
pseudopotentials that I used are robust.

The three bands that lie between $-$7.3 and $-$4.7 eV have a dominant
O $2p$ character, and another six bands with mostly Sb $5p$ character
lie between $-$4.7 and $-$0.5 eV, relative to the Fermi energy. The
bands near the Fermi level have mostly Ti $3d$ character with some
hybridization with Sb $5p$ states, while the Ba $6s$ states are above
the Fermi level. This suggest that Ti ions are nominally trivalent and
are in the $d^1$ state, although the actual electron count will be
different due to some covalency with O $2p$ and Sb $5p$ states.

The Fermi surface (not shown; see Ref.~\onlinecite{sing12}) has a very
two dimensional, square electron sheet around zone corner $M$ with
mixed Ti $d_{z^2}$, $d_{x^2-y^2}$, and $d_{xy}$ characters. There is a
three dimensional hole sheet around $X$ with mostly Ti $d_{z^2}$ with
some admixture of $d_{x^2-y^2}$ and $d_{xy}$ characters. Around the
zone center there is a complex three dimensional electron sheet with
mixed $d_{z^2}$ and $d_{xy}$ character. In the calculations without
spin-orbit coupling, there is also a small electron sheet around the
zone center that has a mixed character. Singh calculated the real part
of the susceptibility within the constant matrix element approximation
and finds peaks in the susceptibility due to substantial nesting
between Fermi sheets, especially near the $\mathbf{q}$ vector
$(\frac{1}{2}, 0)$ (2$\pi/a$). In addition, there is also a smaller
peak near the $\mathbf{q}$ vector $(\frac{1}{2}, \frac{1}{2})$
(2$\pi/a$). This could lead to magnetic instabilities at these
$\mathbf{q}$ vectors, although the height of the peaks could be
diminished when the matrix element is explicitly taken into account in
the calculation of the susceptibility. In any case, Singh finds that
there is a magnetic instability at $X$ that leads to a double stripe
antiferromagnetic order with a doubling of the unit cell along either
the $a$ or $b$ axis with small moments of 0.2 $\mu_B$ per Ti. Based on
the proximity to the SDW instability, Singh predicts a
spin-fluctuation mediated superconductivity in doped BaTi$_2$Sb$_2$O
with a pairing state that has a sign-changing $s$-wave symmetry that
is different from the one in the iron based superconductors.

\begin{figure} %% [tbp]
  \includegraphics[width=\columnwidth]{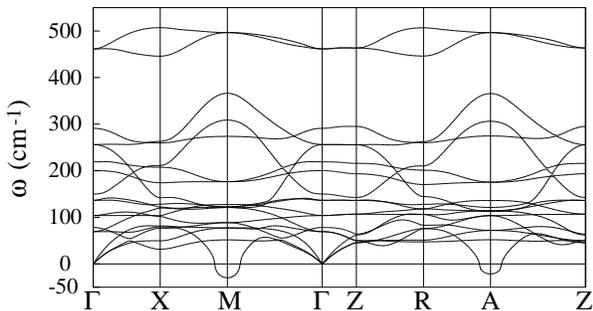}\\
  \caption{Calculated GGA phonon dispersions of non-spin-polarized
    BaTi$_2$Sb$_2$O in the undistorted structure with
    experimental lattice parameters but relaxed internal
    coordinates. The imaginary frequencies are shown as negative.}
  \label{fig:pband}
\end{figure}

\begin{figure} %% [tbp]
  \includegraphics[width=\columnwidth]{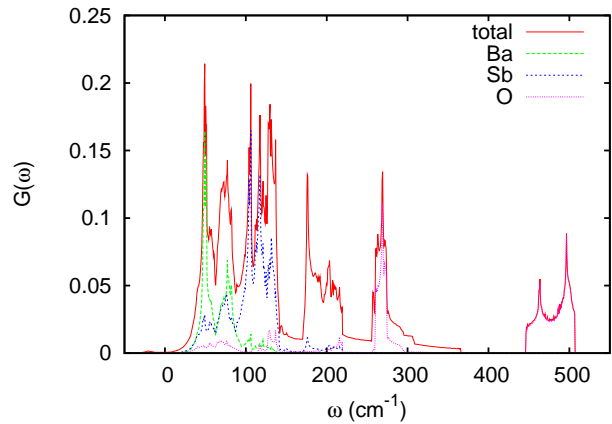}\\\
  \includegraphics[width=\columnwidth]{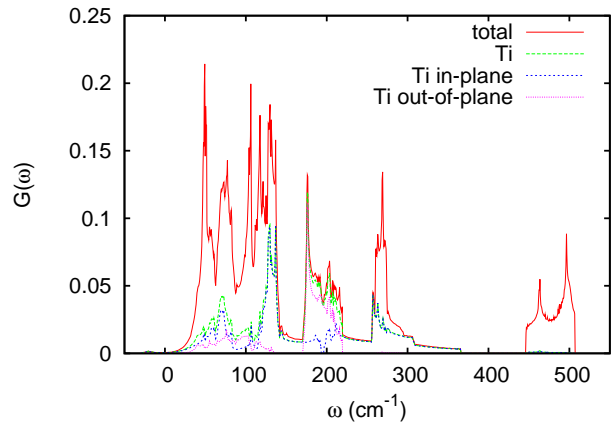}
  \caption{(Color online) Top: Calculated GGA phonon density of states
    $G(\omega)$ along with atomwise projections on Ba, Sb, and O of
    non-spin-polarized BaTi$_2$Sb$_2$O in the undistorted structure
    with experimental lattice parameters but relaxed internal
    coordinates. Bottom: Phonon density of states weighted by
    projections on Ti atom as well as its in-plane and out-of-plane
    characters.}
  \label{fig:pdos}
\end{figure}

\section{Phonons and electron-phonon coupling}

Although there seems to be a magnetic instability in BaTi$_2$Sb$_2$O,
this does not preclude a competing lattice instability or an
electron-phonon superconductivity. In this section, I show results
that indicate that there is also a tendency to a CDW instability,
although the energy of this state is higher than the SDW state
obtained by Singh.\cite{sing12} However, this CDW instability occurs
at the wavevector $(\frac{1}{2}, \frac{1}{2}, 0)$, which is different
from the wavevector $(\frac{1}{2}, 0, 0)$ for the SDW instability. So,
these two instabilities could coexist. Furthermore, I find that the
electron-phonon coupling is large enough to readily explain the
observed superconductivity.

\begin{figure} %% [tbp]
  \includegraphics[width=0.55\columnwidth]{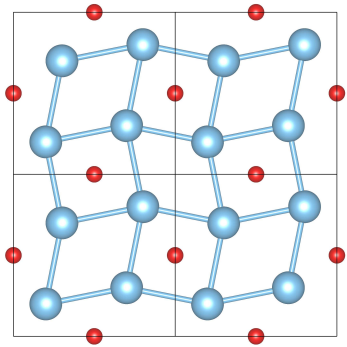}\\
  \caption{(Color online) The distorted Ti plane in the CDW
    phase. The black grid shows the $(\sqrt{2}\times\sqrt{2}\times1)$
    supercell relative to the undistorted structure. The big (cyan)
    balls represent Ti and small (red) balls are O. The Ti
    displacements are exaggerated to make the distortions
    conspicuous.}
  \label{fig:distort}
\end{figure}

\begin{figure} %% [tbp]
  \includegraphics[width=\columnwidth]{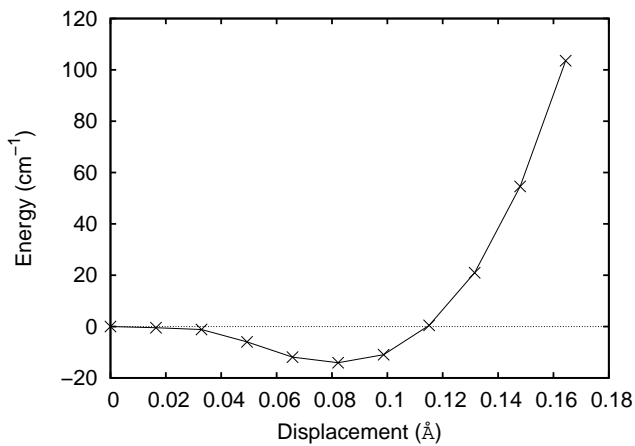}\\
  \caption{Calculated anharmonic double-welled potential for the
    unstable mode obtained using frozen phonon method on a $(\sqrt{2}
    \times \sqrt{2} \times 1)$ supercell. The energy is given per
    formula unit.}
  \label{fig:dblwell}
\end{figure}

The calculated phonon dispersions of BaTi$_2$Sb$_2$O in the
undistorted tetragonal structure with the experimental lattice
parameters and relaxed internal coordinates are shown in
Fig.~\ref{fig:pband}, and the corresponding phonon density of states
along with atomwise projections are shown in Fig.~\ref{fig:pdos}. A
conspicuous feature of the phonon dispersions is the presence of an
unstable mode around $M$ $(\frac{1}{2}, \frac{1}{2}, 0)$ and $A$
$(\frac{1}{2}, \frac{1}{2}, \frac{1}{2})$ indicating a CDW
instability. The instability is weak with the maximum imaginary
frequency of 30$\imath$ cm$^{-1}$, suggesting a very shallow double
well potential. The unstable mode has only in-plane Ti character, and
the distortion corresponds to elongation or compression of the Ti
squares without an enclosed O such that the Ti squares with O rotate
either clockwise or counterclockwise by a small amount (see
Fig.~\ref{fig:distort}).  I also performed relaxation of a $(\sqrt{2}
\times \sqrt{2} \times 1)$ super-cell with the full potential to check
if the instability is a spurious effect of using pseudopotentials, and
I again found a presence of this instability. The Ti ions get
displaced by only 0.096 \AA\ from their original high-symmetry
position. Although the Ti displacements are very small, I find a
substantial reduction of the electronic density of states near the
Fermi level with a reduction from a value of 3.88 eV$^{-1}$ per
formula unit in the undistorted case to a value of 2.83 eV$^{-1}$ per
formula unit in the CDW phase, as obtained from full-potential
calculations without spin-orbit coupling. In addition, I also
performed frozen phonon calculations on the $(\sqrt{2} \times \sqrt{2}
\times 1)$ supercell by displacing Ti atoms according to the
eigenvector of the unstable phonon mode. As shown in
Fig.~\ref{fig:dblwell}, the potential is anharmonic with a minimum at
0.082 \AA, and the depth of the well is very shallow at 14.5
cm$^{-1}$. Hence, it is not surprising that this CDW instability has
so far been eluding detection.

Returning back to the phonon dispersions, the 18 modes of
BaTi$_{2}$Sb$_{2}$O in the undistorted structure extend up to 510
cm$^{-1}$. There is a manifold of 16 bands with mixed Ba, Ti, Sb, and
out-of-plane O character extending up to 370 cm$^{-1}$, which is
separated by a gap from two in-plane O modes between 445 and 510
cm$^{-1}$. Within the lower manifold of 18 bands, the Ba vibrations
lie mostly within 100 cm$^{-1}$. The in-plane Ti character
extends throughout this lower manifold, whereas the out-of-plane Ti
modes are mainly confined to the two narrow bands within 165 and 225
cm$^{-1}$. Most of the Sb weight lies below 145 cm$^{-1}$, and the
out-of-plane O modes lie between 255 and 295 cm$^{-1}$.

\begin{figure} %% [tbp]
  \includegraphics[width=\columnwidth]{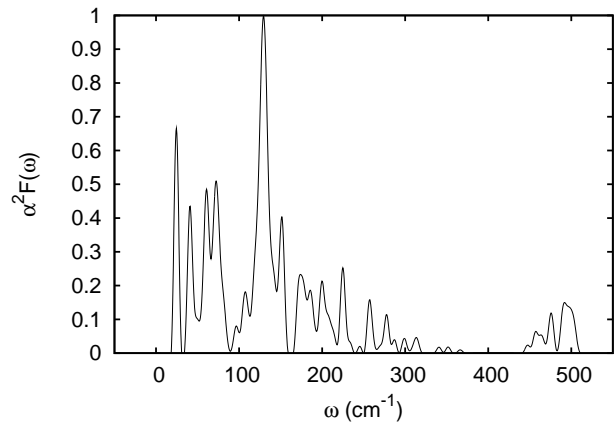}\\
  \caption{Calculated Eliashberg spectral function of BaTi$_2$Sb$_2$O
    in the undistorted structure. The spectral weights for
    imaginary phonon frequencies are set to zero.}
  \label{fig:eliash}
\end{figure}

The strength of the interaction between electrons and phonons is
generally given in terms of the Eliashberg spectral function,
\begin{equation}
\alpha^2 F(\omega)=\frac{1}{N(E_F)}\sum_{\mathbf{k},\mathbf{q},\nu,n,m}%
\delta(\epsilon_{\mathbf{k}}^{n})\delta(\epsilon_{\mathbf{k+q}}^{m}%
)|g_{\mathbf{k},\mathbf{k+q}}^{\nu,n,m}|^{2}\delta(\omega-\omega
_{\nu\mathbf{q}}),
\label{eq:alpha}
\end{equation}
where $N(E_F)$ is the electronic density of states at the Fermi
energy, $\epsilon_{\mathbf{k}}^{n}$ is the electronic energy at
wavevector $\mathbf{k}$ and band index $n$, $\omega _{\nu\mathbf{q}}$
is the energy of a phonon with wavevector $\mathbf{q}$ and branch
index $\nu$, and $|g_{\mathbf{k},\mathbf{k+q}}^{\nu,n,m}|^{2}$ is the
matrix element for an electron in the state $|n\mathbf{k}\rangle$
scattering to $|m\mathbf{k+q}\rangle$ through a phonon $\omega
_{\nu\mathbf{q}}$. The calculated Eliashberg spectral function for
BaTi$_2$Sb$_2$O is shown in Fig.~\ref{fig:eliash}. The spectral weight
is spread through all the phonon frequencies, but it is specially
enhanced at low frequencies below 90 cm$^{-1}$ and between 100 and 160
cm$^{-1}$. I find that a substantial part of the spectral weight comes
from modes that involve in-plane Ti vibrations. This is reasonable as
the states near the Fermi level are dominated by Ti $d$
character. Furthermore, the inter-layer Ti-Ti distance is much larger than
the intra-layer Ti-Ti distance, and the coupling to the out-of-plane Ti
modes is much smaller, as one might expect.

\begin{figure} %% [tbp]
  \includegraphics[width=0.8\columnwidth]{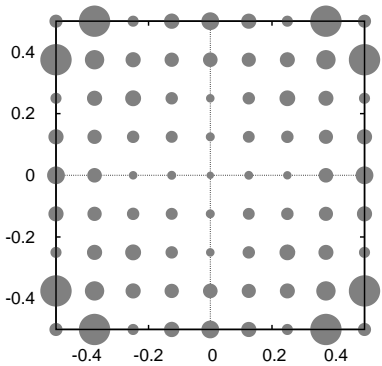}\\
  \caption{Calculated $\mathbf{q}$ dependent total electron-phonon
    coupling $\lambda_{\mathbf{q}}$ shown for the $k_z = 0$ plane of
    the Brillouin zone of the undistorted structure. The area of the
    circles is proportional to the magnitude of
    $\lambda_{\mathbf{q}}$. The unit of the axes is $2\pi/a$.}
  \label{fig:lsumq}
\end{figure}

The $\mathbf{q}$ dependent total electron-phonon coupling
$\lambda_{\mathbf{q}}$ ($= \int_0^{\infty} \mathrm{d}\omega
\alpha^2F(\omega,\mathbf{q})/ \omega$) is plotted for the $k_z = 0$
plane of the Brillouin zone in Fig.~\ref{fig:lsumq}. There are
contributions to the electron-phonon coupling from throughout the
Brillouin zone, but the magnitude of the electron-phonon coupling away
from the zone center is much larger. In particular, the
electron-phonon coupling is peaked near the wavevector $(\pi/a,
\pi/a)$. Again, this is expected from the nesting between the Fermi
sheets near the zone center and zone corners. Furthermore, since the
disconnected Fermi sheets have different orbital character weights,
the electron-phonon coupling on different Fermi sheets are likely to
be different. Hence, this material is very likely to host a multiband
superconductivity. 

The total electron-phonon coupling constant is given by
$\lambda_{\textrm{ep}} \!=\!\sum_{\mathbf{q},\nu}
\lambda_{\mathbf{q},\nu}=2 \int_{0}^{\infty} \mathrm{d}\omega \alpha^2
F(\omega) / \omega$. For BaTi$_2$Sb$_2$O, I obtain
$\lambda_{\textrm{ep}} = 1.28$ with the logarithmically averaged
frequency $\omega_{\ln} = 65$ cm$^{-1}$. (I obtain these numbers by
setting the contribution from the imaginary frequencies to zero.) One
can estimate the $T_c$ within a single-band picture by inserting these
numbers in the simplified Allen-Dynes formula,
\[
T_c= \frac{\omega_{\ln}}{1.2}\exp\left\{ -
\frac{1.04(1+\lambda_{\mathrm{ep}})}{\lambda_{\mathrm{ep}}-\mu^*(1+0.62\lambda_{\mathrm{ep}})}\right\}.
\]
With a value for the Coulomb pseudopotential parameter $\mu^*$ = 0.1,
I obtain $T_c = 9.0$ K, which shows that the conventional
electron-phonon picture readily explains the superconductivity in
BaTi$_2$Sb$_2$O.

The above discussion of the phonon dispersions and the electron-phonon
coupling concerned the undistorted structure, and it showed that
there is both a CDW instability and a strong electron-phonon coupling
that is large enough to yield superconductivity. However, it is more
likely that the electron-phonon superconductivity occurs in the CDW
phase, if it occurs at all. I performed calculations of the phonon
dispersions and electron-phonon coupling for the distorted structure
as well to check if the electron-phonon coupling is strong enough to
give superconductivity.

\begin{figure} %% [tbp]
  \includegraphics[width=\columnwidth]{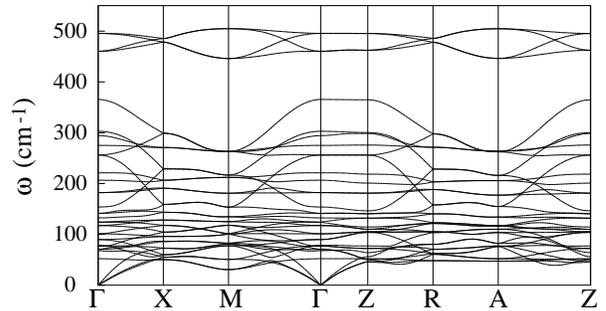}\\
  \caption{Calculated GGA phonon dispersions of non-spin-polarized
    BaTi$_2$Sb$_2$O in the distorted structure.}
  \label{fig:pband-dist}
\end{figure}

\begin{figure} %% [tbp]
  \includegraphics[width=\columnwidth]{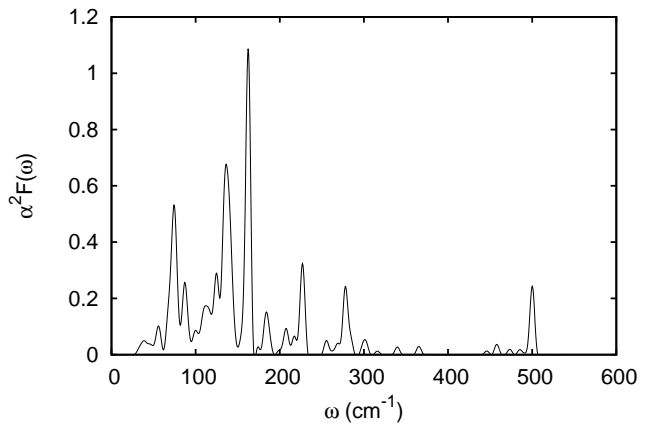}\\
  \caption{Calculated Eliashberg spectral function of BaTi$_2$Sb$_2$O
    in the distorted structure.}
  \label{fig:eliash-dist}
\end{figure}

The phonon dispersions and the Eliashberg spectral function for the
distorted structure are shown in Figs.~\ref{fig:pband-dist} and
\ref{fig:eliash-dist}, respectively. As one expects, the phonon
dispersions do not show any instabilities anymore showing that the
distorted structure is dynamically stable. From
Fig.~\ref{fig:eliash-dist}, we see that the electron-phonon coupling
to the modes below 60 cm$^{-1}$ is greatly reduced as there is no
contribution from the soft in-plane mode that existed in the
undistorted structure. However, the coupling to the modes between 60
and 170 cm$^{-1}$ remains significant. 

In the distorted structure, I find for the total electron-phonon
coupling a value of $\lambda_{\textrm{ep}} = 0.55$ with the
logarithmically averaged frequency $\omega_{\ln} = 110$ cm
$^{-1}$. Using these values in the Allen-Dynes formula as above gives
$T_c$ = 2.7 K, which is comparable to the experimentally obtained
values of 1.2 and 5.5 K.\cite{yaji12,doan12}

%% The electron-phonon calculations indicate that the superconductivity
%% is of multiband strong-coupling nature with the predicted $T_c$ of 9.0
%% K overestimating the experimentally obtained values of 5.5 K and 1.2
%% K.\cite{doan12,yaji12} This discrepancy may be due to the pair
%% breaking effects of spin fluctuations. The total coupling relevant for
%% superconductivity is $\lambda = \lambda_{\textrm{ep}} -
%% \lambda_{\textrm{sf}}$, where $\lambda_{\textrm{sf}}$ is the
%% contribution due to spin fluctuations. The presence of spin
%% fluctuations, as shown to exist in this compound by
%% Singh,\cite{sing12} would therefore reduce the $T_c$.

\section{Conclusions}

In summary, I have presented the results of phonon dispersions and
electron-phonon coupling calculations for BaTi$_2$Sb$_2$O. The phonon
dispersions show a weak lattice instability around Brillouin zone
corners that gives rise to a CDW instability. The distortions
correspond to elongation or compression of the Ti squares without an
enclosed O such that the Ti squares with O rotate either clockwise or
counterclockwise by a small amount. This results in a
$(\sqrt{2}\times\sqrt{2}\times1)$ doubling of the unit cell.

The electron-phonon calculations on the undistorted structure reveal
the presence of strong electron-phonon couplings, especially to the
in-plane Ti modes. There are contributions to the total coupling from
throughout the Brillouin zone, but the magnitude of the couplings are
large away from the zone center. The electron-phonon coupling is
peaked near the zone corners at wavevectors that would correspond to
the nesting vectors between the Fermi sheets around the zone center
and zone corners. The total electron-phonon coupling is
$\lambda_{\textrm{ep}}$ = 1.28, which gives an estimate of $T_c$ = 9.0
K when inserted in the Allen-Dynes formula. The electron-phonon
calculations on the distorted structure also show significant coupling
with a value for the total electron coupling of
$\lambda_{\textrm{ep}}$ = 0.55.  This gives an estimate of $T_c$ = 2.7
K for the distorted structure, which is in good agreement with the
experimentally obtained values.

\section{Acknowledgment}

I gratefully acknowledge the use of the computer cluster of the
Anderson department at Max Planck Institute, Stuttgart, Germany.

\end{document}